\documentclass{article}
\usepackage{spconf,amsmath,graphicx}
\usepackage{epstopdf}
\usepackage{booktabs}
\usepackage{hyperref}
\usepackage{subfigure}
\usepackage{multirow}
\usepackage{makecell}
\renewcommand{\paragraph}[1]{\noindent\textbf{#1}}
\usepackage{fancyhdr}

\title{GI-PIP: Do We Require Impractical Auxiliary Dataset for Gradient Inversion Attacks?}
\name{Yu Sun\textsuperscript{1, \dag},\quad Gaojian Xiong\textsuperscript{1, \dag},\quad Xianxun Yao\textsuperscript{2,*},\quad Kailang Ma\textsuperscript{1},\quad Jian Cui\textsuperscript{1}
\thanks{ \textsuperscript{\dag}These authors have contributed equally to this work} \thanks{ \textsuperscript{*}Corresponding author: yaoxianxun@buaa.edu.cn}} 

\address{\textsuperscript{1}School of Cyber Science and Technology, Beihang University \\
\textsuperscript{2}School of Electronics and Information Engineering, Beihang University}
\begin{document}

%

\maketitle

\thispagestyle{fancy}
\fancyhead{}
\lhead{}
\lfoot{\scriptsize © 2024 IEEE. Personal use of this material is permitted. Permission from IEEE must be obtained for all other uses, in any current or future media, including reprinting/republishing this material for advertising or promotional purposes, creating new collective works, for resale or redistribution to servers or lists, or reuse of any copyrighted component of this work in other works.}
\cfoot{}

\begin{abstract}
Deep gradient inversion attacks expose a serious threat to Federated Learning (FL) by accurately recovering private data from shared gradients. However, the state-of-the-art heavily relies on impractical assumptions to access excessive auxiliary data, which violates the basic data partitioning principle of FL. In this paper, a novel method, Gradient Inversion Attack using Practical Image Prior (GI-PIP), is proposed under a revised threat model. GI-PIP exploits anomaly detection models to capture the underlying distribution from fewer data, while GAN-based methods consume significant more data to synthesize images. The extracted distribution is then leveraged to regulate the attack process as Anomaly Score loss. Experimental results show that GI-PIP achieves a 16.12 dB PSNR recovery using only 3.8\% data of ImageNet, while GAN-based methods necessitate over 70\%. Moreover, GI-PIP exhibits superior capability on distribution generalization compared to GAN-based methods. Our approach significantly alleviates the auxiliary data requirement on both amount and distribution in gradient inversion attacks, hence posing more substantial threat to real-world FL.
Our code is available at \href{https://github.com/D1aoBoomm/GI-PIP}{https://github.com/D1aoBoomm/GI-PIP}.
\end{abstract}

\section{Introduction}

Federated Learning \cite{flancestor} (FL) is a popular distributed learning paradigm that eliminates the direct transmission of training data. Instead, the training process occurs locally on clients, and only shared gradients are sent to a central server for aggregation. Based on the aforementioned design, one has long held the expectation that FL would effectively protect privacy. 

Nevertheless, deep gradient inversion attacks 
\cite{dlg} have revealed that private training data can be accurately recovered by searching the image space for gradient matching. Building upon previous research, numerous attack methods have leveraged prior knowledge to enhance attack performance. For instance, Geiping et al. \cite{ig} have introduced natural image prior and achieved promising attack performance on high-resolution images. Methods based on the generative image prior \cite{gias, ggl} have reduced the image search space to the generator's latent space, achieving state-of-the-art performance.

\begin{figure}[t]
  \centering
  \includegraphics[width=0.95\linewidth]{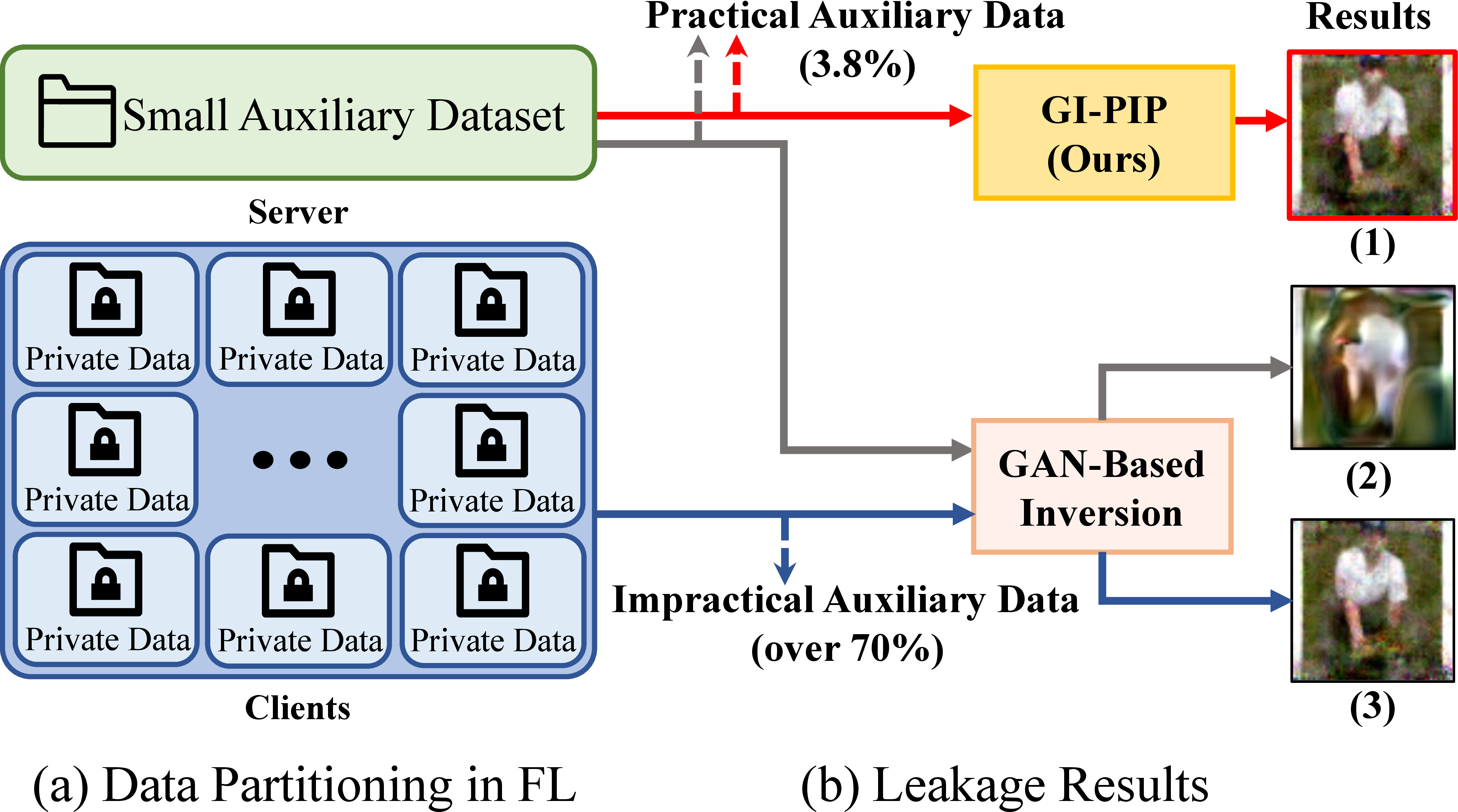}
  \caption{ (a) Data partitioning in FL. (b) Leakage results on ImageNet using various volumes of the auxiliary dataset.}
    \label{concept_figure}
\end{figure}

Although the previous methods \cite{gias,ggl} have effectively utilized prior knowledge to improve attack performance, their threat models are overly restrictive. Specifically, Jeon et al. \cite{gias} and Li et al. \cite{ggl} train GANs on entire training set to conduct attacks, which ignores the basic data partitioning principle of FL, as shown in Figure \ref{concept_figure} (a). In sensitive domains like finance and healthcare, collecting sufficient data proves impractical for adversaries. When only a small dataset is available, severe distortion is observed in images recovered by GAN-based methods, as shown in part (2) of Figure \ref{concept_figure} (b). Therefore, the extent to which gradient inversion compromises privacy in real-world FL settings remains unclear.

To resolve the problems mentioned above, \textit{Gradient Inversion using Practical Image Prior (GI-PIP)} is proposed to improve attack performance under a revised threat model. Within the revised threat model, only a small practical dataset is reasonably held by the honest-but-curious server to maintain the performance of global model. Drawing inspiration from reconstruction-based anomaly detection algorithms \cite{ano_overview,ano_ae1,ano_ae2}, auto-encoders are leveraged to extract the underlying distribution from the practical auxiliary dataset. The extracted prior distribution is subsequently used to regulate the attack as \textit{Anomaly Score (AS) loss}.

Experimental results on image classification datasets indicate GI-PIP's outperformance of existing methods. With only validation set (3.8\% of ImageNet \cite{imagenet}), GI-PIP achieves an impressive PSNR score of 16.12 dB. In contrast, GAN-based methods require impractical 70\% for similar performance. Figure \ref{concept_figure} (b) compares the leakage results of GI-PIP and GAN-based approaches with varying auxiliary data volumes.
Our main contributions can be summarized as follows:

\begin{itemize}
\item The threat model is re-evaluated and standardized (See \S\ref{threat_model_section}), especially regarding the adversary's ability to obtain auxiliary data. This allows for a more realistic assessment of the threat caused by gradient inversion.

\item We propose GI-PIP to perform gradient inversion under the suggested threat model. Anomaly detection models are leveraged to regulate the attack optimization process toward prior distribution (See \S\ref{gi-pip}). 

\item Extensive experiments demonstrate the superiority and effectiveness of GI-PIP (See \S\ref{comparison_section} and \S\ref{Volume_section}). Additionally, GI-PIP exhibits exceptional generalization on distribution compared to existing approaches (See \S\ref{distribution_section}).
\end{itemize}

\section{Methodology}

\subsection{Threat model under basic assumptions} \label{threat_model_section}

An honest-but-curious (HBC) server has been defined as the adversary in previous studies \cite{dlg, ig,gias,ggl,gi,l2i}, but the adversary's capability of collecting auxiliary data is overlooked. In order to resolve this issue, we re-evaluate the threat model and clarify the adversary's capability as follows. 

\begin{figure}[t]
  \centering
  \includegraphics[width=0.95\linewidth]{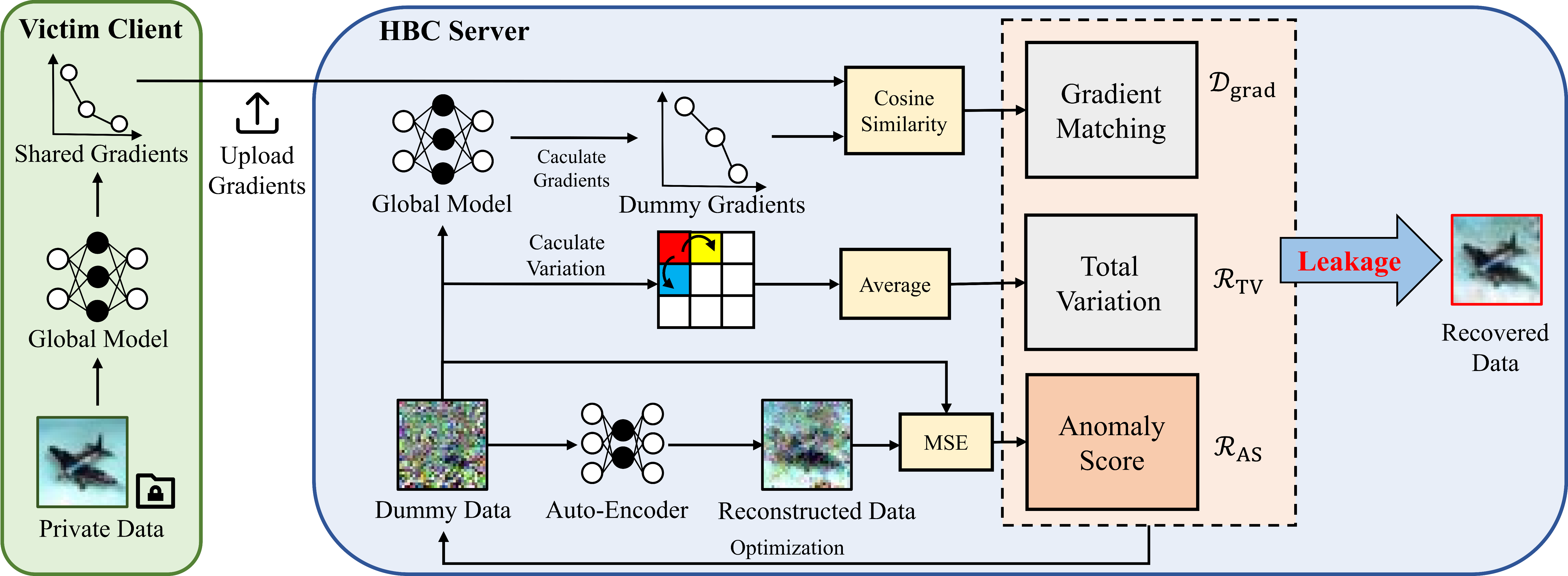}
  \caption{An overview of the GI-PIP framework. On the client side, the victim performs local training normally and transmits shared gradients to the server. On the honest-but-curious server side, under the supervision of the calculated loss, leakage from gradients can be achieved through iterative optimization of the randomly initialized dummy data.}
    \label{method_figure}
\end{figure}

\paragraph{Threat model.} An HBC server is considered the adversary of gradient inversion attacks, as shown in Figure \ref{method_figure}. Merely aiming to recover private data from shared gradients, the server won't behave maliciously (e.g. compromising the models). Inherent access to shared gradients and global model parameters is possessed by the server, which is the fundamental condition for launching gradient inversion attacks. 

\paragraph{Auxiliary dataset.} To maintain the performance of global models, the server is able to hold or collect a small auxiliary dataset reasonably \cite{fl_with_aux, flod, fltrust}. The size of auxiliary dataset is comparable to validation set or test set. The HBC server can conduct gradient inversion attacks using this dataset.


\subsection{Gradient Inversion using Practical Image Prior} \label{gi-pip}

A detailed description of the proposed GI-PIP is presented below. As illustrated in Figure \ref{method_figure}, GI-PIP consists of three main components: Gradient Matching loss, Anomaly Score loss, Total Variation loss. The HBC server randomly initiates the dummy data and computes the loss. Privacy training data will be recovered from shared gradients by iterative optimization.

\paragraph{Objective function.} Balunović et al. \cite{balunovic2021bayesian} formulate the gradient inversion attack problem in a Bayesian framework, which decomposes the recovery of training data into two parts: posterior knowledge provided by gradients and prior knowledge provided by auxiliary data. Given a shared batch-averaged gradient $g$, we consider our optimization problem formally conforms to the Bayesian framework as:

\begin{align}
\hat{x}=\arg \min _{x} \underbrace{\mathcal{R}_\text{AS}(x)+\mathcal{R}_\text{TV}(x)}_{\text{prior knowledge}} + \underbrace{\mathcal{D}_\text{grad}(g^\prime, g)}_{\text{posterior knowledge}},
\end{align}

where $\hat{x} \in R^{N, C, H, W}$ denotes the recovered data and $g^{\prime}=\frac{\partial {Loss}(\mathcal{F}\left(x, \theta_{\mathrm{G}}\right))}{\partial \theta_{\mathrm{G}}}$ represents the dummy gradient generated by dummy data $x$ and global model $\theta_{\mathrm{G}}$. $\mathcal{D}_{\text {grad }}(\cdot)$ is the Gradient Matching loss function. $\mathcal{R}_{\text {AS }}(\cdot)$ and $\mathcal{R}_{\text {TV }}(\cdot)$ respectively represent two regularization terms, AS loss and TV loss. Each of the components is individually explained below.

\paragraph{Anomaly Score loss.} Motivated by reconstruction-based anomaly detection algorithms \cite{ano_ae1,ano_ae2}, we take advantage of auto-encoder models to extract the prior distribution from a small practical dataset. Firstly, anomaly detection models are trained on the test set or validation set to learn to reconstruct the input image. The training process optimizes an auto-encoder model parameterized by $\theta_{\mathrm{A}}$ as follows:

\begin{align}
\min _{\theta_{\mathrm{A}}} \sum_{x_{aux} \in \mathcal{D}_{\text {aux }}}\left\|\mathcal{F}\left(x_{aux}, \theta_{\mathrm{A}}\right)-x_{aux}\right\|^2,
\end{align}

where $\mathcal{F}\left(x_{aux}, \theta_{\mathrm{A}}\right)$ denotes the inference of model, and $x_{aux} \in \mathcal{D}_{\text {aux}}$ means auxiliary data from auxiliary set. A well-trained anomaly detection model is supposed to exactly reconstruct the patches of images that conform to the prior distribution, but is unable to reconstruct anomalous patches well. In gradient inversion, the noisy regions in recovered images are also considered anomalous. Therefore, the Mean Squared Error (MSE) between the recovered image and its reconstructed result is employed to identify the anomalous score of the image. This score acts as a regularization term $\mathcal{R}_{\text {AS }}(x)$ to fix the noise in recovered images. AS loss can be formally described as follows: 

\begin{align} \label{anomaly_loss}
\mathcal{R}_{\text {AS }}(x)=\left\|\mathcal{F}\left(x, \theta_{\mathrm{A}}\right)-x\right\|^2. 
\end{align}

AS loss reflects the degree of conformity between images and the prior distribution. As shown in Equation \ref{anomaly_loss}, anomalous pixels in the image boost the AS loss significantly. By decreasing the AS loss, the anomalous pixels in the image will be corrected, ultimately conforming to the prior distribution, and consequently enhancing the quality of the attack result. In terms of visual effect, the noisy regions will be fixed by the texture of neighboring regions. In \S\ref{ablation_section}, we empirically explain how AS loss regulates the attack through experiments.

\paragraph{Total Variation loss.} Geiping et al. \cite{ig} argue that Total Variation (TV) regularization $\mathcal{R}_{\text {TV }}(x)$ can enhance the recovery effect since it encourages spatial smoothness in images. The TV term is also employed by GI-PIP, which is formulated as:

\begin{align}
\mathcal{R}_{\mathrm{TV}}(x)=\sum_{i, j}\left\|x_{i, j+1}-x_{i, j}\right\|^2+\left\|x_{i+1, j}-x_{i, j}\right\|^2.
\end{align}

This regularization term helps to remove noisy pixels in recovered images, thus enhancing the attack.

\paragraph{Gradient Matching loss.} The Gradient Matching loss provides the major information of training data. This term encourages the dummy data to generate dummy gradients $g^{\prime}$ that are similar to the target gradients $g$. We utilize the Cosine Distance to measure the Gradient Matching loss as follows:

\begin{align}
\mathcal{D}_{\text {grad }}\left(g^{\prime} ; g\right)=1-\frac{<g, g^{\prime}>}{\|g\|_2\left\|g^{\prime}\right\|_2}.
\end{align}

Since Ma et al. \cite{ilrg} have recently demonstrated that training labels for image classification tasks can be perfectly reconstructed in the early stages of training, it is assumed that the labels have been given to generate dummy gradients. 

\section{Experiments}
\subsection{Experimental setup}

\paragraph{FL task.}  Unless otherwise specified, our method is evaluated for image classification tasks on CIFAR-10 dataset \cite{krizhevsky2009learning} and ImageNet dataset \cite{imagenet} (scaled down to $64\times64$ for computational efficiency). A randomly initialized ResNet-18 \cite{resnet} architecture is adopted as the global model.

\paragraph{Data partitioning settings.} Consistent with the capability of the re-evaluated threat model, we train generative models on test or validation set without otherwise specified. All metrics are averaged on 1000 samples from training set.

\paragraph{Evaluation metrics.} Recovered images are provided for visual comparison, and the similarity between the recovered and original training images are quantified by following evaluation metrics: Peak Signal-to-Noise Ratio (PSNR$\uparrow$), Structural Similarity Index Measure (SSIM$\uparrow$) and Learned Perceptual Image Patch Similarity \cite{lpips} (LPIPS$\downarrow$).

\paragraph{Implementation details.} Dummy data are initialized from the standard normal distribution, and are optimized with Adam with 0.1 as learning rate. StyleGAN2 \cite{stylegan} and DCGAN \cite{dcgan} are employed in GIAS. For architecture of AEs or other details, please refer to our code implementation.

\subsection{Comparison with existing methods} \label{comparison_section}

\begin{table}[t]
\centering
\caption{Comparison of GI-PIP and existing methods.}
\smallskip
\resizebox{\linewidth}{!}{
\begin{tabular}{@{}cc|ccc|ccc|ccc@{}}
\toprule
\multirow{2}{*}{Dataset}  & \multirow{2}{*}{Method} & \multicolumn{3}{c|}{Batchsize = 1}      & \multicolumn{3}{c|}{Batchsize = 4}      & \multicolumn{3}{c}{Batchsize = 8}                \\
                          &                         & PSNR↑          & SSIM↑         & LPIPS↓ & PSNR↑          & SSIM↑         & LPIPS↓ & PSNR↑          & SSIM↑         & LPIPS↓          \\ \midrule
\multirow{4}{*}{CIFAR-10} & DLG \cite{dlg}          & 16.85          & 0.61          & 0.4168 & 12.86          & 0.38          & 0.5443 & 11.11          & 0.28          & 0.5489          \\
                          & IG \cite{ig}            & 22.88          & 0.87          & 0.1448 & 15.32          & 0.58          & 0.4157 & 13.04          & 0.40          & 0.5317          \\
                          & GIAS \cite{gias}        & 23.40          & 0.84          & 0.1510 & 16.51          & 0.62          & 0.3983 & 13.44          & 0.41          & 0.5020          \\
                          & \textbf{Ours}           & \textbf{25.49} & \textbf{0.91} & \textbf{0.1271} & \textbf{18.51} & \textbf{0.67} & \textbf{0.3529} & \textbf{16.65} & \textbf{0.57} & \textbf{0.4228} \\ \midrule
\multirow{4}{*}{ImageNet} & DLG \cite{dlg}          & 13.32          & 0.28          & 0.5969 & 12.06          & 0.20          & 0.6505 & 10.15          & 0.16          & 0.6738          \\
                          & IG \cite{ig}            & 16.66          & 0.48          & 0.4939 & 13.69          & 0.27          & 0.6071 & 12.96          & 0.22          & 0.6302          \\
                          & GIAS \cite{gias}        & 15.08          & 0.40          & 0.5333 & 13.36          & 0.28          & 0.5868 & 12.91          & 0.23          & 0.5979          \\
                          & \textbf{Ours}           & \textbf{18.15} & \textbf{0.53} & \textbf{0.4416} & \textbf{16.12} & \textbf{0.41} & \textbf{0.5166} & \textbf{15.54} & \textbf{0.36} & \textbf{0.5334} \\ \bottomrule
\end{tabular}
}
\label{comparison_table}
\end{table}

\begin{figure}[t]
\centering
\includegraphics[width=0.95\linewidth]{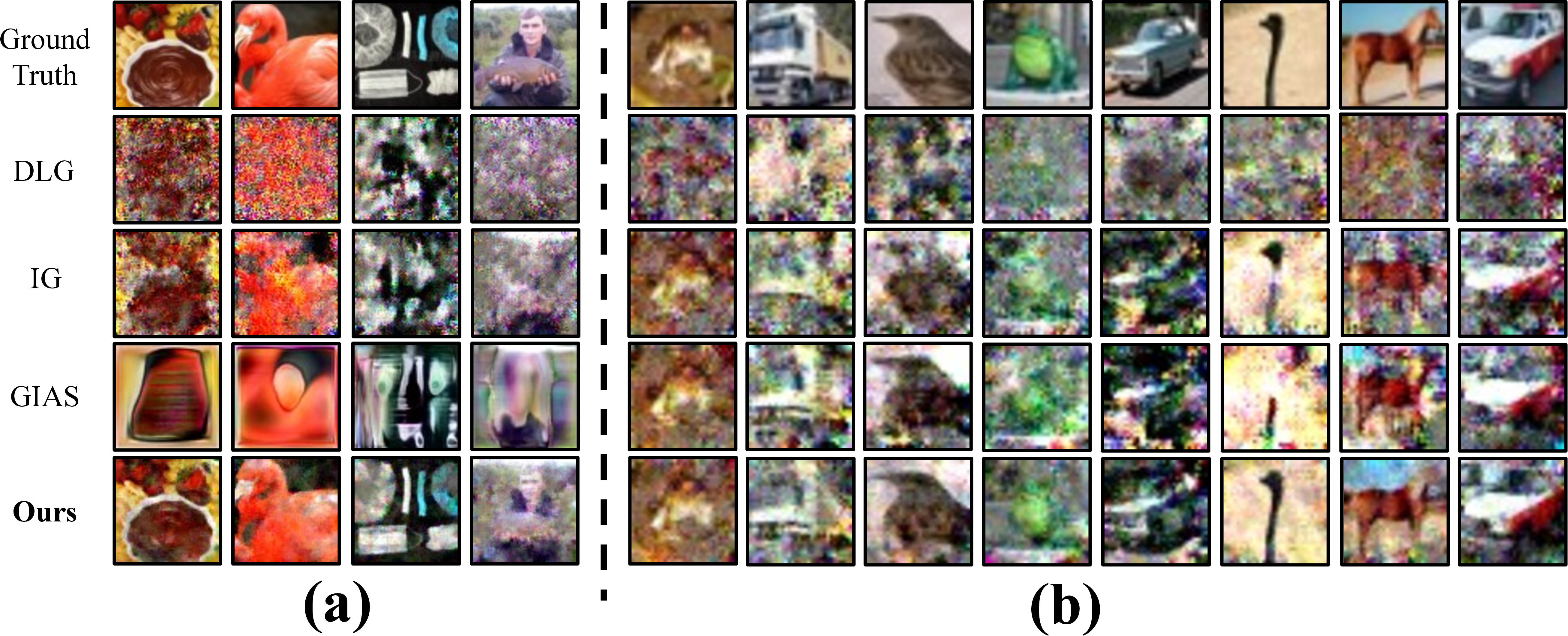}
\caption{Comparison of GI-PIP and existing methods. (a) Batch recovery of ImageNet. (b) Batch recovery of CIFAR10.}
\label{comparison_figure}
\end{figure}

\paragraph{Attack baselines.} GI-PIP is evaluated by comparing it with advanced attack approaches, specifically: (1) Deep Leakage from Gradients (DLG) \cite{dlg}: A method that matches gradients with MSE and leverages LBFGS as optimizer. (2) Inverting Gradients (IG) \cite{ig}: A method that matches gradients using cosine distance loss and optimizes with TV, using Adam as optimizer. (3) Gradient Inversion in Alternative Spaces (GIAS) \cite{gias}: A GAN-based method that matches gradients using latent space search firstly and parameter space search secondly.

\paragraph{Results.} Under the revised threat model, only test set of CIFAR10 (nearly 16.7\%) and validation set of ImageNet (nearly 3.8\%) is used to train GANs and AEs. Table \ref{comparison_table} and Figure \ref{comparison_figure} illustrate the advancement of GI-PIP over prior approaches visually and numerically, respectively. As shown in Figure \ref{comparison_figure}, our results exhibit a remarkable resemblance to the original ones, fully exposing their contours, textures, and other details, thereby posing more serious threat.
As shown in Table \ref{comparison_table}, GI-PIP consistently outperforms existing methods in all the quantitative metrics we considered, including, PSNR, SSIM, and LPIPS. Experimental results demonstrate the effectiveness of the proposed GI-PIP approach.

\begin{figure}[t]
\centering
\includegraphics[width=0.95\linewidth]{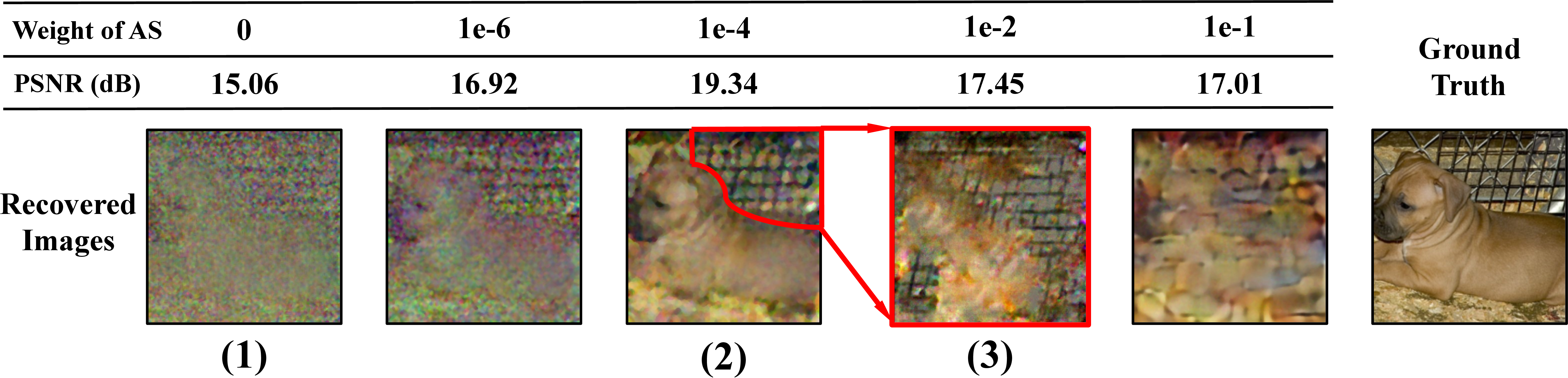}
\caption{Ablation studies on AS loss. For better visual comprehension, a representative sample is chosen and shown.}
\label{ablation_figure}
\end{figure}

\subsection{Ablation studies} \label{ablation_section} To prove the effectivess of AS loss, we conduct single recovery experiments on the high-resolution ImageNet (224$\times$224, for better visualization). We gradually increase the weight of the AS loss and present a series of recovered result in Figure \ref{ablation_figure}. Analysis of the results is as follows: (1) Completely removing AS loss results in a blurry outline. (2) With an increase in the weight of AS loss, the image becomes clearer with refined details, reaching good quality at an appropriate weight of 1e-4. (3) However, when the weight is excessively high, the AS loss \textit{completes the image space using cage texture in the upper-right corner}, leading to the loss of the key subject of the dog. It's shown that AS loss can use texture information from the adjacent area to reconstruct the noisy part of the recovered images, hence enhance attacks.

\begin{figure}[t]
\centering
\includegraphics[width=0.95\linewidth]{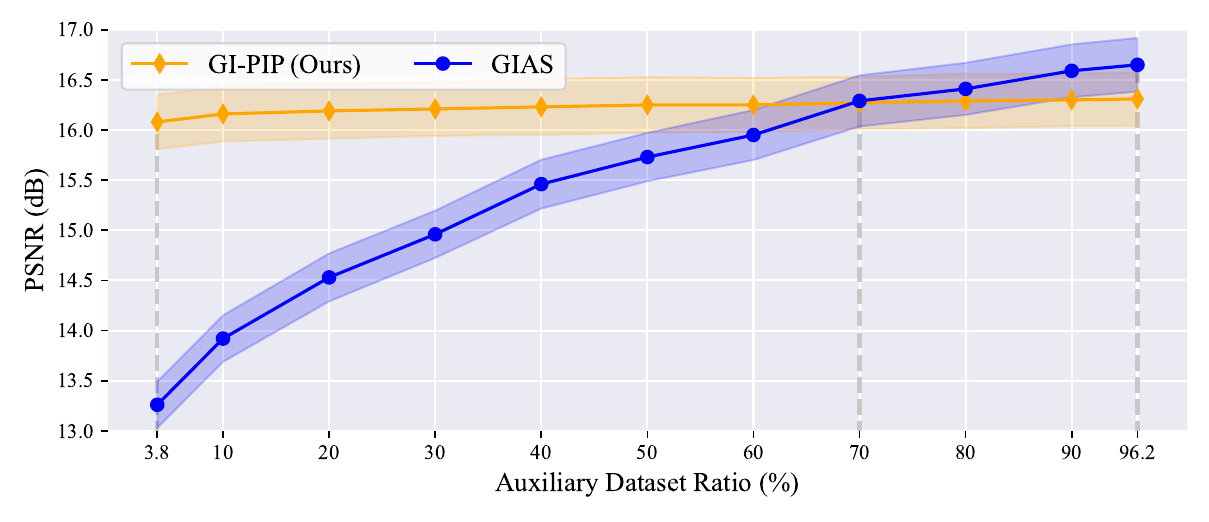}
\caption{Attack performance (with a batch size of 4 on ImageNet) vs volume of auxiliary data. The ratio of auxiliary dataset is gradually increased from $3.8\%$ to $96.2\%$.}
\label{zhexian}
\end{figure}

\subsection{Attacks using varying auxiliary data volume} \label{Volume_section} The volume of auxiliary data is a significant component that affects efficacy of prior knowledge. Our experiment examines how the attack effectiveness of two methods, GI-PIP and GIAS, changes with varying amounts of auxiliary data, as illustrated in Figure \ref{zhexian} (a). The auxiliary data is randomly partitioned from the training set, and the attack targets are chosen randomly from the validation set. It is observed that GI-PIP outperformed GIAS when a small volume of auxiliary data is employed. However, GIAS demonstrates improvement in effectiveness with an increasing amount of auxiliary data, while GI-PIP slightly improves. When the volume of auxiliary data exceeds 70\%, GIAS is proven to be more effective than GI-PIP.
The experimental results demonstrate that GI-PIP does not heavily depend on the amount to improve performance. It must be acknowledged that if excessive auxiliary data is obtained, GIAS remains the most powerful. But when merely practical auxiliary data is available, GI-PIP exhibits the best.

\begin{table}[t]
\centering
\caption{Comparison on distribution generalization ability between GI-PIP and GIAS. Generative models are trained on five animal categories from the test set of CIFAR10.}
\hspace{0pt}
\resizebox{\linewidth}{!}{
\begin{tabular}{@{}ccccc@{}}
\toprule
Target Dataset & Method & PSNR↑ & SSIM↑ & LPIPS↓ \\ \midrule
\multirow{2}{*}{\begin{tabular}[c]{@{}c@{}}In-Distribution\\ (Same Category from CIFAR10 Training Set)\end{tabular}} & GIAS & 16.44 & 0.60 & 0.4059 \\
 & Ours & \textbf{18.62} & \textbf{0.69} & \textbf{0.3423} \\ \midrule
\multirow{2}{*}{\begin{tabular}[c]{@{}c@{}}Similar Distribution\\ (Persian Cat from ImageNet Training Set)\end{tabular}} & GIAS & 13.34 & 0.41 & 0.5058 \\
 & Ours & \textbf{16.88} & \textbf{0.58} & \textbf{0.4207} \\ \midrule
\multirow{2}{*}{\begin{tabular}[c]{@{}c@{}}Out-of-Distribution\\ (Truck Category from CIFAR10 Training Set)\end{tabular}} & GIAS & 9.43 & 0.19 & 0.7332 \\
 & Ours & \textbf{13.12} & \textbf{0.39} & \textbf{0.5120} \\ \bottomrule
\end{tabular}
}
\label{distribution}
\end{table}

\subsection{Distribution generalization capabilities} \label{distribution_section}
Experiments are conducted to compare the distribution generalization capabilities of GI-PIP and existing SOTA (GIAS). Specifically, generative models are trained on the \{bird, cat, dog, horse, deer\} categories from test set of CIFAR10. The distribution similarity is clarified as follows: (1) In distribution: Attack targets are from the same 5 categories in CIFAR10 training set. (2) Similar Distribution: Attack targets are from Persian cat category in ImageNet training set. Considering that CIFAR10 does not feature Persian cats, the distribution of targets is regarded similar but not same. (3) Out-of-distribution: Attack targets are from truck category in CIFAR10 training set, which is completely absent from auxiliary data. As shown in Table \ref{distribution}, when the distribution of target dataset gradually deviates from auxiliary data, both methods exhibit a decline in performance, but GI-PIP consistently performs better. Notably, in out-of-distribution scenarios, the PSNR of GIAS drops to 9.43 dB, while GI-PIP still achieves 13.12 dB. The experimental results fully demonstrate that GI-PIP has better generalization capability than GAN-based methods, hence posing a greater threat to federated learning.

\section{Conclusion}

This study proposes \textit{Gradient Inversion using Practical Image Prior (GI-PIP)}, which utilizes anomaly detection approaches to enhance the attack within the revised threat model. Experimental results demonstrate that GI-PIP effectively compromises the privacy of FL with significantly fewer auxiliary data than SOTA methods. Moreover, the superior capability of the proposed method on distribution generalization is also experimentally indicated. GI-PIP significantly alleviate the auxiliary data requirement on both amount and distribution, bringing gradient inversion closer to more realistic scenarios. We believe our work can contribute towards the development of privacy-preserving FL.

\bibliographystyle{IEEEbib}
\bibliography{main}

\end{document}